\documentclass[nofootinbib,prd,aps]{revtex4}
\usepackage{graphicx}

\begin{document}
\title{Fundamental gravitational limitations to quantum computing}

\author{Rodolfo Gambini$^{1}$, Rafael A. Porto$^{1,2}$ and Jorge Pullin$^{3}$}
\affiliation{1. Instituto de F\'{\i}sica, Facultad de Ciencias,
Igu\'a 4225, esq. Mataojo, Montevideo, Uruguay. \\2. Department of
Physics, Carnegie Mellon University, Pittsburgh, PA 15213\\
 3. Department of Physics and Astronomy, Louisiana State University, 
Baton Rouge, LA 70803-4001}

\date{May 9th 2005}

\begin{abstract}
Lloyd \cite{lloyd} has considered the ultimate limitations physics
places on quantum computers. He concludes in particular that for
an ``ultimate laptop'' (a computer of one liter of volume and one
kilogram of mass) the maximum number of operations per second is
bounded by $10^{51}$.  The limit is derived considering ordinary
quantum mechanics. Here we consider additional limits that are
placed by quantum gravity ideas, namely the use of a relational
notion of time \cite{piombinoprl} and fundamental gravitational
limits \cite{ngwigner} that exist on time measurements. We then
particularize for the case of an ultimate laptop and show that the
maximum number of operations is further constrained to $10^{47}$
per second.

\end{abstract}

\maketitle

Ordinary quantum mechanics is an approximate theory. It is based on  the
existence of a universal and perfectly classical time. Such a notion
is not compatible with our understanding of relativity and
gravitation. The latter requires time to emerge as a
relational variable \cite{PaWo}.  The relation is between the object
being studied and what we choose as a clock to study it. Quantum
mechanics formulated in terms of a relational time differs from
ordinary quantum mechanics \cite{njp}. The ``clock'' is now a quantum
variable with quantum (and thermal) fluctuations. In particular
quantum states lose coherence over time due to the fact that any clock
in nature is imperfect \cite{Bo}. One can find a fundamental level of
loss of coherence that is inescapable if one considers the most
accurate clock that the laws of physics allow us to construct. Such
clocks were considered by Ng and Van Dam \cite{ngwigner}.

What we would like to argue in this paper is that the effect of loss
of coherence due to the lack of a perfect clock in classical mechanics
imposes limitations on the ultimate performance of quantum computers.
Although the effects are derived from gravitation and therefore are
expected to be very small, they are large enough to impose limits on
quantum computation several orders of magnitude smaller than those
found by in reference \cite{lloyd} considering ordinary quantum
mechanics.

In that paper it was noted that for a quantum computer of a finite
energy $E$, one had a finite number of operations that were possible
per second. This is a consequence of the Margolus--Levitin \cite{MaLe}
theorem that states that the time it takes for a quantum state to
evolve into an orthogonal state is $\delta t \ge \pi /(2E)$, where $E$
is the expectation value of the energy and throughout this paper we
choose units so $\hbar=1$.  Therefore a system with energy $E$ can
carry out a maximum of $2E/\pi$ logical operations per second. This
bound is independent on the computer being parallel or serial.
Therefore for a computer weighing one kilogram utilizing all its
mass-energy resources and henceforth operating at the limits of speed
(an ``ultimate laptop''), the maximum number of operations per second
is $n \sim 10^{51}$.

The above reasoning is based on the premise that logical
operations are implemented quantum mechanically by exactly unitary
evolutions. But as we argued above, in quantum mechanics with real
clocks evolution is never perfectly unitary. Therefore there is
the possibility that the above bounds on quantum computation may
have to be revised.

The origin of the lack of unitarity is the fact that definite
statistical predictions are only possible by repeating an
experiment. If one uses a real clock, which has thermal and
quantum fluctuations, each experimental run will correspond to a
different value of the evolution parameter. The statistical
prediction will therefore correspond to an average over several
intervals, and therefore its evolution cannot be unitary. This
effect has been observed in the Rabi oscillations describing the
exchange of excitations between atoms and field \cite{Br}.

There is growing evidence that there exists a fundamental limit to how
accurate a clock can be. Arguments for the limit involve simple
relations derived from basic quantum mechanical principles and
gravitational physics. In their original incarnation the arguments
were based on the fact that more accuracy requires the clock to be
more massive \cite{SaWi} but this in turn is limited in the
gravitational context since large accumulations of mass in small
regions turn the region into a black hole \cite{ngwigner}. This yields
an ultimate accuracy for a clock of the order of $\delta t \ge t_{\rm
  P}^{2/3} t^{1/3}$. where $t_{\rm P}$ is Planck's length. Several
other arguments yield the same limit \cite{lloydng}. Attractively, the
limit also leads naturally to the Bekenstein bound \cite{Ng,Sr}.

The fact that there is a limit to how accurate a clock can be,
coupled with the observation that quantum mechanics with real
clocks fails to be unitary leads naturally to an estimation on the
rate of fundamental loss of coherence of quantum states. In the
limit of highly accurate clocks, the evolution equation of the
density matrix of states takes the approximate form
\cite{njp,piombinoprl},
\begin{equation}
{\partial \rho \over \partial t} = i[\rho,H] -\sigma [[\rho,H],H],
\end{equation}
where $\sigma$ is related to the rate at which the uncertainty in the
clock variable grows. For $\sigma=0$ one would recover ordinary
quantum mechanics. However, as we argued above there is a lower bound
on the value of $\sigma$ motivated from the best possible clock one
can build (a black hole), which turns out to be \cite{piombinoprl},
\begin{equation}
\sigma(t)={ t_P\over 36} \left(\frac{t_P}{T_{\rm
max}-t}\right)^{1/3},
\end{equation}
where $T_{\rm max}$ is the interval of time in which one is interested
in studying the system.


To study the influence of this effect on quantum computers, we start
by recalling the argument by Margolus and Levitin \cite{MaLe} who
showed that to transition from a generic quantum state $|\psi_0>$ to
an orthogonal quantum state $|\psi_t>$ takes a minimum amount of time
$t\ge \pi/(2E)$ where $E$ is the expectation value of energy of the
system (assuming the ground state has zero energy).

As we stated, when one has loss of coherence a state never evolves
completely into an orthogonal one. So if one starts with a density
matrix initially of the form $\rho_{mn}^0=c_m c^*_n$, it will evolve
into
\begin{equation}
\rho^t_{mn} = c_m e^{-i\omega_m t} c^*_n e^{i\omega_n t}
e^{-(\omega_m-\omega_n)^2 t^{4/3}_{\rm P} t^{2/3}},
\end{equation}
and the transition amplitude will be
\begin{equation}
{\rm Tr}\left(\rho^t_{mn} \rho^0_{mn}\right) = \sum_{m,n} |c_m|^2
|c_n|^2 e^{i(\omega_m-\omega_n) t} e^{-(\omega_m-\omega_n)^2
t^{4/3}_{\rm P} t^{2/3}}.
\end{equation}

For instance, let us consider a NOT gate in a quantum computer,
that is a gate that takes a single binary input X and returns the
output 1 if X=0 and 0 if X=1. Taking as initial state
$|\psi_0>=(|E_0>+|E_1>)/\sqrt{2}$ and final state
$|\psi_1>=(|E_0>-|E_1>)/\sqrt{2}$ we will have for the action of the
gate after a time $t\sim \pi/(2E)$,
\begin{equation}
|\psi_0><\psi_0|\rightarrow
(1-\epsilon)|\psi_1><\psi_1|+\epsilon
|\psi_0><\psi_0|,
\end{equation}
with $\epsilon=4 t^{4/3}_{\rm P} t^{2/3} E^2$. Therefore one sees
that the loss of coherence induces error in the quantum
computation and the probability of the error is given by
$\epsilon$ per logical operation.  This effect forces to include
error correction \cite{book}. Error correction also has to be
introduced to compensate for environmental effects, but we are not
considering these here, since we are seeking a fundamental
limit that is inescapable even if one eliminates all environmental
effects. There are fundamental limits on how much error correction
can be introduced in a quantum computer. Although there may be
many mechanisms for error correction, they can all be pictured as
the computer communicating its state to an ``error correcting
device''. Such communication cannot occur faster than the speed of
light.  This limits the rate at which errors can be corrected and
therefore implies a maximum tolerable amount of errors per
operation. The rate at which information can be extracted from a
computer with $L$ bits stored and size $R$ is given by $Lc/R$ and
$c$ is the speed of light. If we divide by the number of
operations per second $n$ we get the maximum tolerable level of
errors per operation for the machine $\epsilon_{\max} \sim
Lc/(nR)$.

The extension of the Margolus--Levitin result to the case in which we
have fundamental decoherence (as we sketched above) establishes a
bound on the speed of operations that is state-dependent. The bound is
saturated when the computer is operating in ``serial mode'', i.e. all
its mass-energy resources and stored bits, $E,L$, are used per logical
operation and therefore is in a state that is a superposition of
states that are widely separated in energy. In such states the
computer achieves a very fast ``clock rate'' $t_{\rm step} \sim 1/E$.
However, it can only carry out a few operations per clock cycle since
its bits are highly entangled. In other words, most of the stored bits
are used to perform a single quantum operation at a high speed. For
these types of states the decoherence effect we are discussing in this
paper is maximum (recall that the effect goes as the energy squared).
On the other hand, if one considers states that are in ``parallel
mode'', that is, where a considerable number of bits are not
entangled and each perform independent quantum operations, the energy
differences are smaller and the decoherence effect gets weaker.
Nonetheless it is still dominant for the case of an ultimate laptop
as we shall see below.

Let us compute the decoherence error one would introduce for a 1kg quantum
computer using a total mass-energy of $E=mc^2 \sim 10^{16} J$ in
serial model. This turns out to be,
\begin{equation}
t_{\rm P}^{4/3} \left({1 \over E}\right)^{2/3} E^2 \sim
10^9\label{serial},
\end{equation}
which is remarkably large. We are therefore led to conclude that such a 
quantum computer cannot utilize all its resources to compute in serial
mode. As it was pointed out by Lloyd \cite{lloyd} an ultimate laptop
working at its maximum of capacity would have a degree of
parallelization ($d_p$), defined in Ref.~\cite{lloyd} to be roughly
$d_p \sim 1/{\epsilon}_{max}$, of the order of $d_p \sim 10^{10}$ for
the ultimate laptop. It is therefore crucial to extend our effect to
the case of parallel computation. It is easy to see that the only
difference with the calculation in (\ref{serial}) is that now the
energy is redistributed amongst $L/d_p$ parallel qubits and therefore
the energy per gate goes down to $E/d_p$ (the case $d_p=1$ will
account naturally for the serial mode). Similarly to what happens in
the serial case, we will be also led to conclude that an ultimate
laptop cannot utilize all its resources without running into an
unavoidable error crash. In order to see this let us particularize the
bound we previously obtained for the error rate for the case of a
quantum computer of size $R$ and $L$ bits stored,
\begin{equation}
t_{\rm P}^{4/3} \left({d_p \over E_{\rm eff}}\right)^{2/3}
\left(\frac{E_{\rm eff}}{d_p}\right)^2 \le {c L\over n R},
\end{equation}
where $E_{\rm eff}$ is the {\it effective} energy the quantum computer
can actually invest with a degree of parallelization $d_p$. Given
now that the number of operations per second in the later is
bounded by $n < E_{\rm eff}$, we have that,
\begin{equation}
n\le \left({1\over t_{\rm P}}\right)^{4/7} \left({cL \over
R}\right)^{3/7}d_p^{4/7} \sim 10^{47} {\rm op/s}\label{n}.
\end{equation}

This expression is general for a quantum computer of $L$ bits and
characteristic size $R$ operating with degree of parallelization
$d_p$, and the numerical estimate is obtained from Ref.~\cite{lloyd}
where $ L \sim 10^{31}$, $R \sim 0.1 m$ and $d_p \sim 10^{10}$ for an
ultimate laptop with volume one liter (this is also related to the
maximum entropy that can be contained in the volume \cite{Be}). In
addition, (\ref{n}) also leads us to conclude that a 1 kg quantum
computer working at serial mode ($d_p=1$) can not perform more than
$10^{42}$ op/s.

These bounds, though large, are three and nine orders of magnitude
more stringent for parallel and serial computation respectively, than
the one found by Lloyd (that yields the same bound for both cases).

Finally, if one is interested in miniaturization, one may wish to consider what
are the limits on the most compact computer one can manufacture. Such
a computer is a black hole, as argued by Ng and Lloyd
\cite{Ngprl,lloydng}. In this case the maximum number of bits is given
by the formula of Bekenstein. A similar calculation to the one above
leads to an estimate of $n\le (M/M_{\rm Planck})^{3/7}/t_{\rm Planck}$
that for a kilogram mass black hole is approximately $10^{47} {\rm
  op/s}$. So the black hole computer faces the same limitations as the
``ultimate laptop'' due to the effect we consider.

Finally, let us add that if one wishes to consider an ``Avogadro
computer'', a more realistic sort of computer in which qubits consist
of atomic nuclei, the maximum number of operations per second is
reduced to $10^{39}$ working in serial mode (this is largely due to
the fact that the number of qubits $L$ that appears in the above
formulae is reduced to $L\sim 10^{25})$, a bound slightly tighter
than the one implied by the Margolus--Levitin theorem ($10^{40}$).
If the computer operates in parallel one can reach the latter limit.

Summarizing, we have found that formulating quantum mechanics properly
in terms of realistic clocks yields fundamental limitations on
quantum computers that are more stringent than other bounds of
similar nature found up to present.

We wish to thank Jack Ng for useful comments. This work was
supported by grant NSF-PHY0090091, NASA-NAG5-13430 and funds from
the Horace Hearne Jr. Laboratory for Theoretical Physics and
CCT-LSU. The work of R.A.P. is supported in part by the Department
of Energy under grants DOE-ER-40682-143 and DEAC02-6CH03000.

\end{document}